\documentclass[11pt]{article}
\usepackage{exscale}\usepackage{graphicx}
\def\d{\mbox{\rm d}}
\begin{document}
\title{On the inverse problem of calculus of variations}
 \author{M.C. Nucci  $\;$ and $\;$ A.M. Arthurs\footnote{Permanent address: Department of
Mathematics, University of York, Heslington, York YO10 5DD, U.K.,
e-mail: ama5@york.ac.uk}}
\date{Dipartimento di Matematica
e Informatica, Universit\`a di Perugia, 06123 Perugia, Italy,
e-mail: nucci@unipg.it}

 \maketitle
 \begin{abstract}
  We show that given an ordinary differential equation of  order four, it may be possible
  to determine a Lagrangian if the third derivative is absent (or eliminated)
  from the equation. This represents a subcase of Fels'conditions
  [M. E. Fels, The inverse problem of the calculus of variations
for scalar fourth-order ordinary differential equations, { Trans.
Amer. Math. Soc.} { 348} (1996) 5007-5029] which ensure the
existence and uniqueness of the Lagrangian in the case of a
fourth-order equation.
 The key is the Jacobi last multiplier as in the case of a second-order equation.
 Two equations from a Number Theory paper
by Hall,
  one of second and one of fourth order,  will be used to exemplify the
  method. The known link between Jacobi last multiplier and Lie
symmetries is also exploited. Finally the Lagrangian of two
fourth-order equations drawn from Physics are determined with the
same method.
 \end{abstract}
\section{Introduction}

It is well-known that a Lagrangian always exists for any
second-order ordinary differential equation \cite{Whittaker}. What
 seems less known is that the key is the Jacobi last multiplier
\cite{Jacobi 44 b}, \cite{Jacobi 45 a}, \cite{Jacobi 86 a},
\cite{Whittaker}, which has many interesting properties, a list of
which can be found in \cite{jlm05}.

It is a matter for historians alike to find out why Darboux
\cite{Darboux 94}, Helmholtz \cite{Helm 87}, Koenigsberger
himself\footnote{In 1902-1903 Koenigsberger wrote  Helmholtz's
biography \cite{Koenig 03} -- which in 1906 was (abridged)
translated into English  with a Preface by Lord Kelvin \cite{Welby
06} -- after he wrote his 1901-book on Mechanics \cite{Koenig 01}.
Neither books cite the connection between Jacobi last multiplier
and Lagrangians. In 1904 Koenigsberger wrote Jacobi's biography
\cite{Koenig 04} where the Jacobi Last Multiplier is extensively
described.} and many other successive authors, e.g. Douglas
\cite{Douglas 41} and Havas\footnote{Havas even cites the book by
Whittaker \cite{Whittaker} but only in connection with the
formulation of Lagrangian equations.} \cite{Havas 57}, never
acknowledged the use of the Jacobi last multiplier in order to
find Lagrangians of a second-order equation.

The Norwegian Sophus Lie, who carefully studied Jacobi's work
\cite{Hawkins 91}, found a connection between his groups of
transformations and the Jacobi last multiplier \cite{Lie1874},
\cite{Lie 12 a}. The Italian Bianchi presented Lie's work in his
lectures on finite continuous groups
 of transformations, and described quite clearly
 the Jacobi Last multiplier and its properties \cite{Bianchi 18}.
 Neither of them cited the connection
 with Lagrangians.

A deluge of papers have been dedicated to the solution of the
inverse problem of calculus of variation, namely finding a
Lagrangian of differential equations.
 Most of them did not acknowledge  the seminal work of Jacobi. Some did (see references in
 \cite{jlm05}, and  \cite{CP07Rao1JMP}).

In \cite{Fels96} Fels derived the necessary and sufficient
conditions under which a fourth-order equation,
 i.e.
\begin{equation}
u^{(iv)}=F(t,u,u',u'',u'''). \label{geno4}
\end{equation} admits a unique
Lagrangian, namely:
\begin{eqnarray}
\frac{\partial^3 F}{\partial (u''')^3}=0\label{Fels1}\\
\frac{ \partial F}{\partial u'} + \frac{1}{2}\frac{{\rm d}^2}{{\rm
d} t^2}\left(\frac{ \partial F}{\partial u'''}\right) - \frac{{\rm
d}}{{\rm d} t}\left(\frac{ \partial F}{\partial u''}\right) -
\frac{3}{4} \frac{ \partial F}{\partial u'''} \frac{{\rm d}}{{\rm
d} t}\left(\frac{
\partial F}{\partial u'''}\right) + \frac{1}{2} \frac{ \partial
F}{\partial u''} \frac{ \partial F}{\partial u'''} + \frac{1}{8}
\left( \frac{
\partial F}{\partial u'''} \right)^3 = 0.\label{Fels2}
\end{eqnarray}
 Here we propose to extend the use of the Jacobi last multiplier
in order to find the Lagrangian for ordinary differential
equations of order four satisfying a subset of Fels' conditions
(\ref{Fels1})-(\ref{Fels2}).

The paper is organized in the following way. In the next section
we  present the properties of the Jacobi last multiplier and its
connection to Lie symmetries; then we show the connection of the
Jacobi last multiplier with Lagrangians for any second-order
equation, and how and when can this connection be extended to
fourth-order equations. In section 3 two equations from a Number
Theory paper by Hall \cite{Hall1},
  one of second and one of fourth order,  are used to exemplify the
  method. The link between Jacobi last multiplier and Lie
symmetries \cite{Lie1874}, \cite{Lie 12 a} is also exploited. In
section 4 we consider two fourth-order equations drawn
  from Physics and use the method of the Jacobi last multiplier to find their respective
   Lagrangians, and lastly we
  provide some of the many examples where the method  does not work.
   The last section contains some final remarks.

\section{Jacobi Last Multiplier}
The method of the Jacobi Last Multiplier \cite{Jacobi 44
b}-\cite{Jacobi 86 a}) provides a means to determine all the
solutions of the partial differential equation
\begin {equation}
\mathcal{A}f = \sum_{i = 1} ^n a_i(x_1,\dots,x_n)\frac {\partial
f} {\partial x_i} = 0 \label {2.1}
\end {equation}
or its equivalent associated Lagrange's system
\begin {equation}
\frac {\d x_1} {a_1} = \frac {\d x_2} {a_2} = \ldots = \frac {\d
x_n} {a_n}.\label {2.2}
\end {equation}
In fact, if one knows the Jacobi Last Multiplier and all but one
of the solutions, then the last solution can be obtained by a
quadrature. The Jacobi Last Multiplier $M$ is given by
\begin {equation}
\frac {\partial (f,\omega_1,\omega_2,\ldots,\omega_{n- 1})}
{\partial (x_1,x_2,\ldots,x_n)}
 = M\mathcal{A}f, \label {2.3}
\end {equation}
where
\begin {equation}
\frac {\partial (f,\omega_1,\omega_2,\ldots,\omega_{n- 1})}
{\partial (x_1,x_2,\ldots,x_n)} = \mbox {\rm det}\left [
\begin {array} {ccc}
\displaystyle {\frac {\partial f} {\partial x_1}} &\cdots &\displaystyle {\frac {\partial f} {\partial x_n}}\\
\displaystyle {\frac {\partial\omega_1} {\partial x_1}} & &\displaystyle {\frac {\partial\omega_1} {\partial x_n}}\\
\vdots & &\vdots\\
\displaystyle {\frac {\partial\omega_{n- 1}} {\partial x_1}}
&\cdots &\displaystyle {\frac {\partial\omega_{n- 1}} {\partial
x_n}}
\end {array}\right] = 0 \label {2.4}
\end {equation}
and $\omega_1,\ldots,\omega_{n- 1} $ are $n- 1 $  solutions of
(\ref {2.1}) or, equivalently, first integrals of (\ref {2.2})
independent of each other. This means that  $M$ is a function of
the variables $(x_1,\ldots,x_n)$ and  depends on the chosen $n-1$
solutions, in the sense that it varies as they vary. The essential
properties of the Jacobi Last Multiplier are:
\begin{description}
\item{ (a)} If one selects a different set of $n-1$ independent
solutions $\eta_1,\ldots,\eta_{n-1}$ of equation (\ref {2.1}),
then the corresponding last multiplier $N$ is linked to $M$ by the
relationship:
$$
N=M\frac{\partial(\eta_1,\ldots,\eta_{n-1})}{\partial(\omega_1,
\ldots,\omega_{n-1})}.
$$
\item{ (b)} Given a non-singular transformation of variables
$$
\tau:\quad(x_1,x_2,\ldots,x_n)\longrightarrow(x'_1,x'_2,\ldots,x'_n),
$$
\noindent then the last multiplier $M'$ of  $\mathcal{A'}F=0$ is
given by:
$$
M'=M\frac{\partial(x_1,x_2,\ldots,x_n)}{\partial(x'_1,x'_2,\ldots,x'_n)},
$$
where $M$ obviously comes from the $n-1$ solutions of
$\mathcal{A}F=0$ which correspond to those chosen for
$\mathcal{A'}F=0$ through the inverse transformation $\tau^{-1}$.
\item{ (c) } One can prove that each multiplier $M$ is a solution
of the following
 linear partial differential equation: \begin {equation}
\sum_{i = 1} ^n \frac {\partial (Ma_i)} {\partial x_i} = 0; \label
{2.5} \end {equation} \noindent viceversa every solution $M$ of
this equation is a Jacobi Last Multiplier.
\item{ (d) } If one
knows two Jacobi Last Multipliers $M_1$ and $M_2$ of equation
(\ref {2.1}), then their ratio is a solution $\omega$ of (\ref
{2.1}), or, equivalently,  a first integral of (\ref {2.2}).
Naturally the ratio may be quite trivial, namely a constant.
Viceversa the product of a multiplier $M_1$ times any solution
$\omega$ yields another last multiplier
$M_2=M_1\omega$.\end{description}
  Since the existence of a solution/first integral
is consequent upon the existence of symmetry, an alternate
formulation in terms of symmetries was provided by Lie
\cite{Lie1874}, \cite {Lie 12 a}. A clear treatment of the
formulation in terms of solutions/first integrals  and symmetries
 is given by Bianchi \cite {Bianchi 18}. If we know
$n- 1 $ symmetries of (\ref {2.1})/(\ref {2.2}), say
\begin {equation}
\Gamma_i =
\sum_{j=1}^{n}\xi_{ij}(x_1,\dots,x_n)\partial_{x_j},\quad i = 1,n-
1, \label {2.6}
\end {equation}
Jacobi's last multiplier is given by $M =\Delta ^ {- 1} $,
provided that $\Delta\not = 0 $, where
\begin {equation}
\Delta = \mbox {\rm det}\left [
\begin {array} {ccc}
a_1 &\cdots & a_n\\
\xi_{1,1} & &\xi_{1,n}\\
\vdots & &\vdots\\
\xi_{n- 1,1}&\cdots &\xi_{n- 1,n}
\end {array}\right]. \label {2.8}
\end {equation}
There is an obvious corollary to the results of Jacobi mentioned
above. In the case that there exists a constant multiplier, the
determinant is a first integral.  This result is potentially very
useful in the search for first integrals of systems of ordinary
differential equations.  In particular, if each component of the
vector field of the equation of motion is missing the variable
associated with that component, i.e., $\partial a_i/\partial x_i =
0 $, the last multiplier is a constant, and any other Jacobi Last
Multiplier is a first integral.

Another property of the Jacobi Last Multiplier is  its (almost
forgotten) relationship with the Lagrangian, $L=L(t,u,u')$, for
any second-order equation
\begin{equation}
u''=F(t,u,u') \label{geno2}
\end{equation}
namely \cite{Whittaker}:
\begin{equation}
M=\frac{\partial^2 L}{\partial u'^2} \label{relMLo2}
\end{equation}
where $M=M(t,u,u')$ satisfies the following equation
\begin{equation} \frac{{\rm d}M}{{\rm d} t}+M\frac{\partial F}{\partial
u'} =0.\label{Meqo2}
\end{equation}
Then equation (\ref{geno2}) becomes the Euler-Lagrange equation:
\begin{equation}
-\frac{{\rm d}}{{\rm d} t}\left(\frac{\partial L}{\partial
u'}\right)+\frac{\partial L}{\partial u}=0. \label{ELo2}
\end{equation}
The proof is given by taking the derivative of (\ref{ELo2}) by
$u'$ and showing that this yields (\ref{Meqo2}).
% i.e. df(EL,ux)=eqm;
 If one knows a Jacobi last multiplier, then $L$ can be
  obtained by a double integration, i.e.:
\begin{equation}
L=\int\left (\int M\, {\rm d} u'\right)\, {\rm d}
u'+f_1(t,u)u'+f_2(t,u),
\end{equation}
where $f_1$ and $f_2$ are functions of $t$ and $u$ which have to
satisfy a single partial differential equation related to
(\ref{geno2}) \cite{laggal}. As it was shown in \cite{laggal},
$f_1, f_2$ are related to the gauge function $g=g(t,u,u')$. In
fact, we may assume
\begin{eqnarray}
f_1&=&  \frac{\partial g}{\partial u}\nonumber\\
f_2&=& \frac{\partial g}{\partial t} +f_3(t,u) \label{gf1f2o2}
\end{eqnarray}
where $f_3$ has to satisfy the mentioned partial differential
equation and $g$ is obviously arbitrary. The importance of the
gauge function should be stressed. In order to apply Noether's
theorem correctly, one should not assume $g\equiv const$,
otherwise some first integrals may not be found (see \cite{laggal}
and the second-order equation in the next section).

We now consider a fourth-order equation (\ref{geno4}). In this
case the Jacobi last multiplier satisfies the following equation
\begin{equation} \frac{{\rm d}M}{{\rm d} t}+M\frac{\partial F}{\partial
u'''} =0.\label{Meqo4}
\end{equation}
It is trivial to show from Fels'conditions
(\ref{Fels1})-(\ref{Fels2}) that if a Lagrangian $L=L(t,u,u',u'')$
is taken such that
\begin{equation}
M=\frac{\partial^2 L}{\partial u''^2} \label{relMLo4}
\end{equation}
along with the constraints\footnote{We note that (\ref{vinco4_2})
resembles  the Euler-Lagrange equation (\ref{ELo2}) with
$F,u',u''$ replacing $L,u,u'$ respectively.}
\begin{eqnarray}
\frac{\partial F}{\partial u'''} =0, \label{vinco4_1} \\\frac{
\partial F}{\partial u'} - \frac{{\rm
d}}{{\rm d} t}\left(\frac{
\partial F}{\partial u''}\right)  = 0 \label{vinco4_2}
\end{eqnarray}
then equation (\ref{geno4}) becomes the Euler-Lagrange equation:
\begin{equation}
\frac{{\rm d}^{2}}{{\rm d}t^{2}}\left(\frac{\partial L}{\partial
u''}\right)-\frac{{\rm d}}{{\rm d} t}\left(\frac{\partial
L}{\partial u'}\right)+\frac{\partial L}{\partial u}=0.
\label{ELo4}
\end{equation}
We emphasize  that because of the assumption (\ref{vinco4_1}),
then a Jacobi last multiplier $M$ is trivial to find from equation
(\ref{Meqo4}), namely
\begin{equation} M={\rm const}.\end{equation} Then $L$ can be   obtained by a double integration,
i.e.:
\begin{equation}
L=\frac{M}{2}u''^2+f_1(t,u,u')u''+f_2(t,u,u'), \label{Lo4}
\end{equation}
where $f_1$ and $f_2$ are functions of $t,u,u'$ which have to
satisfy some partial differential equations related to
(\ref{geno4}). We can   relate $f_1, f_2$ to the gauge function
$g=g(t,u,u')$. In fact, we may assume
\begin{eqnarray}
f_1&=&  \frac{\partial g}{\partial u'}\nonumber\\
f_2&=& \frac{\partial g}{\partial u}u'+\frac{\partial g}{\partial
t} +f_3(t,u,u') \label{gf1f2o4}
\end{eqnarray}
where $f_3$ has to satisfy the mentioned partial differential
equations  and $g$ is obviously arbitrary. Again we stress the
importance of the gauge function. In order to apply Noether's
theorem correctly, one should not assume $g\equiv const$,
otherwise some first integrals may not be found as in the examples
of the next section.

\section{Two examples from Number Theory}
In this section we show in details the application of the Jacobi
Last multiplier along with its connection to Lie symmetries. We
also solve the inverse problem of calculus of variations as given
in \cite{Hall1}.
\subsection{A second-order equation}
In \cite{Hall1} the following functional was introduced:
\begin{equation}
\int_0^{\pi} y'^4+6\nu y^2y'^2 d\,x,\label{6}
\end{equation}
where $y=y(x)\in C^2[0,\pi]$, $y(0)=y(\pi)=0$ and $\nu\geq 0$. The
corresponding Euler-Lagrange equation is
\begin{equation}
y'^2 y''+\nu y^2y''+\nu y y'^2=0\label{eq6}\,. \end{equation} If
we apply Lie group analysis to this equation, we find that it
admits a two-dimensional abelian transitive Lie symmetry algebra
(Type I) generated by the following two operators:
\begin{equation}
\Gamma_1=\partial_x,\;\;\;\; \Gamma_2=y\partial_y \,.\label{op6}
\end{equation}
Then we can   integrate equation (\ref{eq6}). First, we introduce
a basis of differential invariants of $\Gamma_1$, i.e.:
\begin{equation}
 u=\frac{y'}{y},\;\;\; v=x\,. \label{inv6}\end{equation}
 Then equation (\ref{eq6}) reduces to the following first-order equation:
 \begin{equation}\frac{{\rm d}u}{{\rm d}v}=\frac{ - \nu u v}{\nu v^2 +
 u^2},\end{equation}
 which admits the operator $\Gamma_2$ in the space of variables $u,v$, i.e.
 \begin{equation}\Gamma_2=v\partial_{v}+u\partial_{u}\,.\end{equation}
 Then its general solution is implicitly given by:
 \begin{equation}
 \sqrt{u}\left(2\nu v^2 + u^2\right)^{1/4}=const
 \end{equation}
 and in the original variables\footnote{The same first integral can be obtained by using
  Noether's theorem (see below)}:
 \begin{equation}
 \sqrt{y'}\left(2\nu y^2 + y'^2\right)^{1/4}=const \label{i1}
 \end{equation}
 viz
 \begin{equation}
 y'=
 \frac{\sqrt{\left(\nu
 y^2a_1+\sqrt{\nu^2y^4a_1^2+1}\right)
 y^2a_1}}{ya_1\left(\nu y^2a_1+\sqrt{\nu^2y^4a_1^2+1}\right)}
       \end{equation}
       with $a_1$ an arbitrary constant
 %y(0) implies y'(0)=1/\sqrt{a_1}, (see hall1-6.mws)
Finally the general solution of (\ref{eq6}) is given implicitly
by:
\begin{equation}
 \int \frac{ya_1\left(\nu
y^2a_1+\sqrt{\nu^2y^4a_1^2+1}\right)}{\sqrt{\left(\nu
 y^2a_1+\sqrt{\nu^2y^4a_1^2+1}\right)
 y^2a_1}}\,d\,y=x+a_2
\end{equation}
We  note that if $y$ is positive and $a_1=1$ the integral on the
left-hand side could be integrated in terms of an hypergeometric
function $\cal H$, namely:
\begin{equation}
\frac{1}{2}\sqrt{2\nu}y^2{\cal H}\left(\left[-\frac{1}{2},
\frac{1}{4}, -\frac{1}{4}\right],\left[\frac{1}{2},
\frac{1}{2}\right],-\frac{1}{y^4\nu^2}\right)
\end{equation}
Let us try to find a Lagrangian for equation (\ref{eq6}) by using
the Jacobi last multiplier, namely through  (\ref{relMLo2}). The
two Lie point symmetries (\ref{op6}) yield a Jacobi last
multiplier. In fact the following matrix \cite{Lie1874}, \cite{Lie
12 a}:
\begin {equation}
\left (\begin {array} {ccc} 1& y'&- {\displaystyle{\frac{\nu yy'^2}\nu y^2+y'^2}}\\
 1& 0& 0\\
 0& y& y'
\end {array}
\right)
\end {equation}
has determinant different from zero and its inverse is a Jacobi
last multiplier, i.e.
\begin{equation}
M_1=-\frac{2\nu y^2+y'^2}{y'^2(\nu y^2+y'^2)}. \label{M1eq6}
\end{equation}
The corresponding Lagrangian is
\begin {equation}
L_1=\frac{1}{4\nu y}\left(
-\sqrt{2\nu}y'\arctan\left(\frac{y'}{\sqrt{2\nu}y}\right)+
\log(2\nu y^2 + y'^2)\nu y
        + 2\log(y')\nu y \right)+ f_1(x,y)y' + f_2(x,y), \label{L1}
        \end {equation}
where $f_1, f_2$ are solutions of
 \begin {equation}
        \frac{\partial f_1}{\partial x}-\frac{\partial f_2}{\partial
        y}=0. \label{eqf1f2}
\end{equation}
If we impose the link between $f_1, f_2$ with the gauge function
$g(x,y)$, namely (\ref{gf1f2o2}), then $f_3(x,y)$ becomes just
$f_3(x)$, an arbitrary function of the independent variable $x$.
The Lagrangian (\ref{L1}) may appear ugly. Nevertheless the
corresponding variational problem admits two Noether's symmetries,
namely both Lie symmetries given in (\ref{op6}) are Noether's
symmetries. Consequently the following two first integrals of
equation (\ref{eq6}) can be   found by applying Noether's
theorem\footnote{Also the corresponding gauge function $g$ is
given.  It is important to remark that in the case of the first
integral (\ref{int2}) the gauge function $g$ cannot be constant,
while it can be constant in the case of the first integral
(\ref{int1}). Naturally, we have left out any inessential additive
constants.}:
\begin{eqnarray}
\Gamma_1 &\Rightarrow&  I_1=(2\nu y^2 + y'^2)y'^2, \quad\quad
\left[g=e^{x}\left(\int\frac{f_3(x)}{e^x}\,{\rm d}x
 + a_2\right)\right] \label{int1}\\
\Gamma_2 &\Rightarrow& I_2=\frac{1}{4\nu y'}\left( -
\sqrt{2\nu}y'\arctan\left(\frac{y'}{\sqrt{2\nu}y}\right) + 2\nu y-
4\nu x y'\right), \quad[g= s(x)y + x] \label{int2}.
\end{eqnarray}
with $s(x)$ an arbitrary function of $x$. We note that the first
integral $I_1$ in (\ref{int1}) was already derived in
(\ref{i1}).\\
At this point one would like to know if it is possible to obtain
the original Lagrangian given in (\ref{6}), i.e.
\begin{equation}
L_H=y'^4+6\nu y^2y'^2
\end{equation}
 A property of the Jacobi last
multiplier is that if one knows a Jacobi last multiplier and a
first integral then their product gives another multiplier
\cite{jlm05}. If we take the product of the first integral $I_1$
(\ref{int1}) and the multiplier $M_1$ (\ref{M1eq6}), then we
obtain another Jacobi last multiplier of equation (\ref{eq6}),
i.e.:
\begin{equation}
M_2=-y'^2-\nu y^2
\end{equation}
which can be integrated twice with respect to $y'$ in order to
yield the following Lagrangian\footnote{Note the inessential
multiplicative constant.}:
\begin{equation}
L_2=-\frac{1}{12}(y'^4+6\nu y^2y'^2)+f_1y'+f_2,
\end{equation}
where $f_1, f_2$ are solutions of (\ref{eqf1f2}).  It is
interesting to emphasize that this Lagrangian (namely Hall's
Lagrangian) is such that the Lie operator $\Gamma_2$ in
(\ref{op6}) does not generate a Noether's symmetry for the
corresponding variational problem. In fact only
$\Gamma_1$  is the generator of a Noether's symmetry for Hall's Lagrangian.\\

 If instead of (\ref{6}) we consider the functional
\cite{Hall1}
\begin{equation}
\int_0^{\pi} y'^4+6\nu y^2y'^2 -3\lambda(\nu) y^4\,d\,x.\label{6c}
\end{equation} and apply Lie group analysis to its corresponding  Euler-Lagrange equation, viz:
\begin{equation}
y'^2 y''+\nu y^2y''+\nu y y'^2+\lambda y^3=0\label{eq6c}\,,
\end{equation}
then we obtain the same Lie symmetry algebra generated by
(\ref{op6}) which means that (\ref{eq6c}) can be integrated by
quadrature. In fact if we introduce the same variables as in
(\ref{inv6}) then equation (\ref{eq6c}) reduces to the following
first-order equation:
 \begin{equation}
 \frac{{\rm d}u}{{\rm d}v}=\frac{ -  v\left(\lambda v^2+\nu u^2\right)}
 {u\left(\nu v^2 + u^2\right)},\end{equation}
which can  be integrated to give
\begin{equation}
\lambda v^4 + 2\nu u^2 v^2 + u^4=const
\end{equation}
and in the original variables
\begin{equation}
\lambda y^4 + 2\nu y'^2 y^2 + y'^4=const
\end{equation}
a first integral of equation (\ref{eq6c}).

 Also Lie group
analysis implies that if $\lambda=\nu^2$ then equation
(\ref{eq6c}) admits an eight-dimensional Lie symmetry algebra
generated by the following operators:
\begin{eqnarray}
\Gamma_1&=&-y\left(-\cos(\sqrt{\nu}x)\partial_x+y\sqrt{\nu}\sin(\sqrt{\nu}x)\partial_y\right)\nonumber\\
\Gamma_2&=&-y\left(\sin(\sqrt{\nu}x)\partial_x+y\sqrt{\nu}\cos(\sqrt{\nu}x)\partial_y\right)\nonumber\\
\Gamma_3&=&\cos(2\sqrt{\nu}x)\partial_x-y\sqrt{\nu}\sin(2\sqrt{\nu}x)\partial_y\nonumber\\
\Gamma_4&=&-\sin(2\sqrt{\nu}x)\partial_x-y\sqrt{\nu}\cos(2\sqrt{\nu}x)\partial_y\nonumber\\
\Gamma_5&=&\partial_x\\
\Gamma_6&=&y\partial_y\nonumber\\
\Gamma_7&=&\cos(\sqrt{\nu}x)\partial_y\nonumber\\
\Gamma_8&=&-\sin(\sqrt{\nu}x)\partial_y.\nonumber
\end{eqnarray}
  which means that equation
(\ref{eq6c}) is linearizable or indeed linear. Indeed in this case
equation (\ref{eq6c})  is just \begin{equation}y''=-\nu y.
\label{linear6}\end{equation} In order to find Lagrangians and
first integrals for equation (\ref{6c}) we have to repeat mutatis
mutandis the same analysis given above. We may underline  that in
the case of $\lambda=\nu^2$, namely equation (\ref{linear6}), we
can generate 14 different Lagrangians as it was shown in
\cite{laggal}.

\subsection{A fourth-order equation}
Another functional in \cite{Hall1} is the following:
\begin{equation}
\int_0^\pi y'^4+\mu y^2y''^2\,d\,x \,.\label{calcvareq26}
\end{equation}
The corresponding Euler-Lagrange equation is
\begin{equation}
4\mu y y' y'''+\mu y^2y^{iv}+2\mu y'^2y''+3\mu
yy''^2-6y'^2y''=0\,.\label{eq26}
\end{equation}
 If we apply Lie group analysis to
this equation, we find that it admits a three-dimensional  Lie
symmetry algebra
 generated by the following three operators:
\begin{equation}
X_1=\partial_x,\;\;\;\; X_2=y\partial_y, \;\;\;\;
X_3=x\partial_x\,,\label{op26}
\end{equation}
which means that we can reduce equation (\ref{eq26}) to a
first-order equation, i.e.: \begin{equation} \frac{{\rm d}\tilde
u}{{\rm d}\tilde x}=\frac{ - 7\mu \tilde u \tilde x - \mu\tilde u
- 6\mu \tilde x^3 - 4\mu \tilde x^2 + 6\tilde x}{\mu\tilde
u}\end{equation} with \begin{equation}\tilde
u=\frac{y'''y^2}{y'^3}-2\frac{y''^2y^2}{y'^4}+\frac{y''y}{y'^2},\;\;\;\;
\tilde x=\frac{y''y}{y'^2}\,.
\end{equation}
 If $\mu=3$ then equation (\ref{eq26}) admits
an eight-dimensional Lie symmetry algebra $\cal{L}$ generated by
the following eight operators:
\begin{eqnarray}
\Lambda_1= x^2\partial_x+\frac{3}{2}x y\partial_y,\quad \quad
\Lambda_2=x\partial_x, \quad\quad\Lambda_3=\partial_x, \quad\quad
\Lambda_4=y\partial_y, \quad\quad
\Lambda_5=\frac{x^3}{y}\,\partial_y,\nonumber \\
\Lambda_6=\frac{x^2}{y}\,\partial_y, \quad\quad
\Lambda_7=\frac{x}{y}\,\partial_y, \quad\quad
\Lambda_8=\frac{1}{y}\,\partial_y \label{Lambdas}
\end{eqnarray}
 This means that equation (\ref{eq26}), i.e.:
\begin{equation}
4  y' y'''+y y^{iv}+3y''^2=0 \label{eq26lin}
\end{equation}
  is
linearizable by means of a point transformation \cite{Lie 12 a}.
In order to find the linearizable transformation we have to find
an abelian intransitive two-dimensional subalgebra of $\cal{L}$
and, following Lie's classification of two-dimensional algebras in
the real plane \cite{Lie 12 a}, we have to transform it into the
canonical form
\begin{equation}\partial_{ u},\;\;\;\;\;\ t\partial_{u}\label{Icanform}\end{equation}
 with $ u$ and $t$
the new dependent
 and independent variables, respectively.
We found that one such subalgebra
  is that generated by  $\Lambda_7$ and $\Lambda _8$. Then we have to solve the
following four linear partial differential equations of first
order:
\begin{equation}
\Lambda_7(t) = 0,\quad \Lambda_8(t) = 0,\quad \Lambda_7(u) =
 t, \quad \Lambda_8(u) = 1.
\end{equation}
It is readily shown that the  linearizable transformation is
\begin{equation}t=x,\quad\quad\quad u=y^2\label{trlino4}, \end{equation}
 and equation (\ref{eq26lin}) becomes
\begin{equation} u^{iv}=0.\label{ux40}\end{equation} Finally, the general solution of (\ref{eq26lin})  is
\begin{equation} y=\sqrt{a_1 + a_2 x + a_3 x^2 + a_4
x^3}\end{equation} with $a_i (i=1,4)$ arbitrary constants.\\ We
note that if we apply the transformation (\ref{trlino4}) to
equation (\ref{eq26}) in the case of any $\mu$ then  the following
equation is obtained:
\begin{equation}
u^{iv}=- \frac{(\mu - 3)(2uu'' - u'^2)u'^2}{4\mu u^3}
\label{eq26m}
\end{equation}
which does not contain $u'''$, namely the third derivative of $u$
by $x$. Therefore a constant, say 1, is a Jacobi last multiplier
of equation (\ref{eq26m}), and we can   obtain a Lagrangian from
(\ref{Lo4}), i.e.:
\begin{equation}
L_1=\frac{1}{2}u''^2+f_1(x,u,u')u''+f_2(x,u,u')\label{Leq26m}
\end{equation}
where $f_1$ and $f_2$ satisfy (\ref{gf1f2o4}) and $f_3$ has to
have the following expression:
\begin{equation}
f_3(x,u,u')=\frac{( - \mu + 3)u'^4}{24\mu u^2}+h_1(x,u)u'+
h_2(x,u)
\end{equation}
with $h_1, h_2$ arbitrary functions of $x,u$.

 We remark that a
Jacobi last multiplier of equation (\ref {eq26}) can be derived
from equation (\ref{Meqo4}), i.e.:
\begin{equation}
\frac{1}{M}\frac{{\rm d} M}{{\rm d} x} -4 \frac{y'}{y}=0 \quad
\Longrightarrow \quad M=y^4, \label{jlmeqo4}
\end{equation}
although this Jacobi last multiplier is useless in order to find a
Lagrangian of equation (\ref {eq26}). Instead, if we apply
transformation (\ref{trlino4})  to (\ref{Leq26m}) in order to go
back to the original function $y(x)$, then adding also the
particular assumptions $f_1=-u'^2/(2 u)$, and $f_2=u'^4(\mu +
1)/(16u^2)$ yield the following Lagrangian for equation
(\ref{eq26}) :
\begin {equation}
Lm_1=(\mu-1) y'^4 + 2y^2y''^2
\end {equation}
which, apart from an inessential multiplicative constant, is
Hall's Lagrangian in (\ref{calcvareq26}).
\\
Let us apply Noether's theorem. If we consider the Lagrangian in
(\ref{Leq26m}), we find that the following two first integrals of
equation (\ref{eq26m}) can be obtained:
\begin{eqnarray}
\frac{3}{2}X_3+X_1 &\Rightarrow&  I_1= \frac{1}{8\mu u^2}\, (12\mu
u^3 u''' - 4\mu u^2  u'  u'' - 8  \mu  u^2  u'  u''' x+ 4  \mu
u^2  u''^2  x \nonumber \\&&\quad\quad + 2  \mu  u  u'^3-
\mu  u'^4  x - 6  u  u'^3 + 3  u'^4  x), \nonumber\\
X_2 &\Rightarrow& I_2=\frac{1}{8\mu u^2} (- 8\mu u^2 u'u''' + 4\mu
u^2 u''^2 - \mu u'^4 + 3 u'^4 ).  \label{i2eq26m}
\end{eqnarray}
A similar result is obtained if one uses Hall's Lagrangian and
equation (\ref{eq26}). Moreover, if we apply Noether's theorem to
the linearizable  equation (\ref{eq26lin}) with Hall's Lagrangian,
i.e.:
\begin{equation}
L_H=y'^4+3y^2y''^2,
\end{equation}
we obtain the following seven first integrals\footnote{Naturally,
they are not all independent from each other.}:
\begin{eqnarray}
\Lambda_1 &\Rightarrow&  Im_1=- 3 y^3 y'' + 3 y^3 y''' x + y^2
y'^2 + 7 y^2 y' y'' x \nonumber \\&&\quad\quad- 2 y^2  y' y''' x^2
+ y^2 y''^2 x^2 - 2 y y'^3 x - 4 y y'^2 y'' x^2 +
y'^4 x^2 \nonumber\\
\frac{3}{4}\Lambda_4+\Lambda_2
 &\Rightarrow& Im_2=9 y^3 y''' + 21 y^2 y' y'' - 12 y^2 y' y''' x + 6 y^2
y''^2 x - 6 y y'^3 - 24 y y'^2 y'' x + 6 y'^4 x\nonumber\\
\Lambda_3 &\Rightarrow& Im_3=- 2 y^2 y' y''' + y^2 y''^2 - 4 y y'^2 y'' + y'^4 \nonumber\\
\Lambda_5 &\Rightarrow& Im_5=- 3 y^2 + 6 y y' x - 3 y y'' x^2 + y
y''' x^3 - 3 y'^2 x^2 + 3
y' y'' x^3 \nonumber\\
\Lambda_6 &\Rightarrow& Im_6=2 y y' - 2 y y'' x + y y''' x^2 - 2 y'^2 x + 3 y' y'' x^2 \nonumber\\
\Lambda_7 &\Rightarrow& Im_7=- y y'' + y y''' x - y'^2 + 3 y' y'' x \nonumber\\
\Lambda_8 &\Rightarrow& Im_8=y y''' + 3 y' y''\label{inteq26lin}
\end{eqnarray}

Although we do not write down the corresponding expressions  of
the gauge function, we underline that it cannot always be  set
equal to a constant otherwise none of $Im_1, Im_5, Im_6, Im_7$,
and $Im_8$ could be obtained.

All seven first integrals (and even more) may be obtained without
Noether's Theorem. In fact we just need to find the Jacobi last
multipliers of equation (\ref{eq26lin}) that are obtained by
inverting the nonzero determinants of the possible 70 matrices
made out of the eight Lie symmetries (\ref{Lambdas}). Then the
ratio of any two multipliers is a first integral of equation
(\ref{eq26lin}). For example:
\begin {equation}
C_{1234}=\left (\begin {array} {ccccc} 1& y'& y''& y'''&
-{\displaystyle{\frac{4y'y'''+3y''^2}{y}}}\\ [0.3cm] x^2 &
\frac{3}{2}y x & \frac{3}{2} y-\frac{1}{2} x y' & y'-\frac{5}{2} x
y'' &
-\frac{3}{2} y''-\frac{9}{2} x y'''\\ [0.2cm]  x & 0 & -y'& -2y''& -3y'''\\
 1& 0& 0&0& 0\\
 0& y& y'& y''& y'''
\end {array}
\right)
\end {equation}
 is
the matrix obtained by considering the symmetries generated by
operators $\Lambda_1, \Lambda_2, \Lambda_3, \Lambda_4$ in
(\ref{Lambdas}); its determinant is
\begin{equation}
\Delta_{1234}=9 y'' y y' y'''-2 y'^3 y'''-\frac{3}{2} y'^2 y''^2+6
y''^3 y+\frac{9}{2} y'''^2 y^2,
\end{equation}
and the corresponding Jacobi last multiplier is:
\begin{equation}
M_{1234}=\frac{1}{\Delta_{1234}}=\frac{1}{18 y'' y y' y'''-4 y'^3
y'''-3 y'^2 y''^2+12 y''^3 y+9 y'''^2 y^2}.
\end{equation}
Similarly, the matrix $C_{5678}$ yields the determinant
$\Delta_{5678}=12/y^4$, i.e. the Jacobi last
multiplier\footnote{The multiplicative constant is inessential.}:
\begin{equation}
M_{5678}=y^4
\end{equation}
which we have already found in (\ref{jlmeqo4}) as an obvious
solution of (\ref{Meqo4}). Note that
\begin{equation}
\frac{M_{5678}}{M_{1234}}=y^4(18 y'' y y' y'''-4 y'^3 y'''-3 y'^2
y''^2+12 y''^3 y+9 y'''^2 y^2)
\end{equation}
is ``another" first integral of equation (\ref{eq26lin}).

\section{More fourth-order equations}

A similarity reduction of the fifth-order Korteweg-De Vries
equation is the following fourth-order equation \cite{Cosgrove
00}, \cite{Lit99}
\begin {equation}
u^{iv}=20u u''+10 u'^2-40 u^3+\alpha u+t. \label{litteq}
\end{equation}
This equation satisfies Fels' conditions
(\ref{Fels1})-(\ref{Fels2}) as well as
(\ref{vinco4_1})-(\ref{vinco4_2}). Therefore $M=1$ is a Jacobi
last multiplier, and formula (\ref{relMLo4}) yields the following
Lagrangian of (\ref{litteq}):
\begin {equation}
L= \frac{1}{2}(u''^2+t^2u')+ \alpha t u u' - 40 t u^3 u' + 10 u
u'^2 +\frac{{\rm d}}{{\rm d}t} g(t,u,u')
\end{equation}
where $g=g(t,u,u')$ is the gauge function.

In a paper on  wave caustics \cite{Kit94}, Kitaev derived the
following equation:
\begin {eqnarray}
u^{iv}&=&\frac{1}{8u^2}( 16 u u' u'''- 12 \alpha u^5 + 8 \beta u^3
- 20 u^7 + 40 u^4 u'' + 8 u^4 t + 20 u^3  u'^2   - 8 u u'\nonumber
\\ &&
+ 12 u u''^2 - 8 u u'' t + u y^2 - 16  u'^2 u'' + 8 u'^2 t)
\label{kiteq}
\end{eqnarray}
which satisfies Fels' conditions (\ref{Fels1})-(\ref{Fels2}), and
admits $M=u^{-2}$ as a Jacobi last multiplier.
 The simple transformation $u=w^2$ transforms equation
(\ref{kiteq}) into the following equation in $w=w(t)$
\begin {eqnarray}
w^{iv}&=&\frac{1}{16w^3}( - 12 \alpha w^8 + 8 \beta w^4 - 20 w^12
+ 80 w^7 w'' + 160 w^6 w'^2+ 8 w^6 t
     + 160 w w'^2 w'' \nonumber \\ &&- 16 w w' - 16 w w'' t - 80 w'^4 + 16 w'^2 t +
     t^2)\label{kiteqw}
     \end{eqnarray}
    This equation satisfies Fels' conditions
(\ref{Fels1})-(\ref{Fels2}) as well as
(\ref{vinco4_1})-(\ref{vinco4_2}). Therefore $M=1$ is a Jacobi
last multiplier, and formula (\ref{relMLo4}) yields the following
Lagrangian for equation (\ref{kiteqw}):
\begin {eqnarray}
L&=& \frac{1}{2}w''^2+\frac{w'}{48 w^3}( - 36 \alpha w^8 t + 24
\beta w^4 t - 60 w^{12} t + 120 w^7 w'
    + 12 w^6 t^2 + 40 w w'^3 \nonumber \\ &&- 24 w w' t + t^3) +\frac{{\rm d}}{{\rm d}t} g(t,w,w')
\end{eqnarray}
where $g=g(t,w,w')$ is the gauge function. Replacing $w=\sqrt{u}$
into this Lagrangian yields the Lagrangian for equation
(\ref{kiteq}), i.e.:
\begin {eqnarray}
L&=& \frac{1}{96u^3} ( 12 u^2 u''^2 - 12 u u'^2 u'' - 36 \alpha
u^5 u' t + 24 \beta u^3 u' t - 60 u^7 u' t + 60 u^4 u'^2
   + 12 u^4 u' t^2 \nonumber \\ && - 12 u u'^2 t + u u' t^3
      +8 u'^4)+\frac{{\rm d}}{{\rm d}t} g(t,u,u')
\end{eqnarray}
where $g=g(t,u,u')$ is the gauge function.

The method we proposed here is not the ultimate method. We can
list many equations for which it does not work. In
\cite{marcicnuc} seven fourth-order equations were derived as
similarity reductions of a mathematical model  for thin liquid
films \cite{jain} and its corresponding heir equations
\cite{itera}. None of the seven equations passes condition
(\ref{Fels2}), although they all satisfy condition (\ref{Fels1}).

\section{Final remarks}

When one deals with a second-order differential equation the
following remarks should be enlightened and kept in mind:
\begin{itemize}
\item The most efficient Lagrangian, namely that which allows the
most number of Noether's symmetries, may not be  the Lagrangian
with the  simplest form.
\item Lie symmetries are the key tool for finding Jacobi last multipliers
 and therefore Lagrangians.
\end{itemize}

In \cite{Fels96}  the necessary and sufficient conditions under
which a fourth-order equation (\ref{geno4}) admits a unique
Lagrangian were determined. In this paper we suggest that the
application of the Jacobi last multiplier may yield that unique
Lagrangian in some instances.

If one rewrites a fourth-order equation as either a suitable
system of two second-order equations \cite{Douglas 41}, or a
system of four first-order equations \cite{Havas 73} the challenge
of solving an inverse problem of calculus of variations may still
be open.

%Even more if nonlocal variables are involved.

\section*{Acknowledgements}

AMA is indebted to: Professor Gianluca Vinti for arranging his
visits to the Dipartimento di Matematica e Informatica,
Universit\`a di Perugia; Professor Bryce McLeod for initiating the
connection in the first place; and finally to them and their
colleagues, Professors Anna Rita Sambucini and Carlo Bardaro, for
excellent hospitality during his visits. The support of the
Erasmus Exchange scheme is gratefully acknowledged.

 \end{document}